\UseRawInputEncoding
\documentclass[aps,prb,showpacs,notitlepage,floatfix,superscriptaddress,
twocolumn]{revtex4-1}
\usepackage{bbm}
\usepackage{mathrsfs}
\usepackage{epsfig}
\usepackage{graphicx}
\usepackage{amsfonts}
\usepackage[figuresright]{rotating}
\usepackage{amssymb}
\usepackage{amsmath}
\usepackage{dcolumn}
\usepackage{bm}
\usepackage{braket}
\usepackage[colorlinks,linkcolor=blue,anchorcolor=blue,citecolor=blue,urlcolor=blue]{hyperref}
\usepackage{multirow}
\usepackage{longtable}
\usepackage{pbox}
\usepackage{float}

\def\be{\begin{equation}} \def\ee{\end{equation}}
\def\bp{\begin{pmatrix}} \def\ep{\end{pmatrix}}
\def\bea{\begin{eqnarray}}
\def\eea{\end{eqnarray}}
\def\beaa{\begin{equation}\begin{aligned}}
\def\eeaa{\end{aligned}\end{equation}}

\def\nn{\nonumber}

\begin{document}
\title{An electromagnetic way to derive basic relativistic transformations}
\author{Congjun Wu}
\affiliation{Department of Physics, School of Science, Westlake University, Hangzhou 310024, Zhejiang, China}
\affiliation{
Institute for Theoretical Sciences, Westlake University, Hangzhou 310024, Zhejiang, China}
\affiliation{Key Laboratory for Quantum Materials of Zhejiang Province, School of Science, Westlake University, Hangzhou 310024, Zhejiang, China}
\affiliation{Institute of Natural Sciences, Westlake Institute
for Advanced Study, Hangzhou 310024, Zhejiang, China}

\begin{abstract}
We derive the relativistic velocity addition law, the transformations of electromagnetic fields and space-time intervals by examining
the drift velocities in a crossed electromagnetic field configuration.
The postulate of light velocity invariance is not taken
as a priori, but is derived as the universal upper limit of
drift velocities.
The key is that a physical drift motion of either an electric charge
or a magnetic charge remains a drift motion by any inertial
reference frame transformations.
Such a simple fact is incompatible with the Galilean
velocity addition.
This derivation provides a way to introduce relativity via elementary
electromagnetism.
\end{abstract}
\maketitle

As is well known that the discovery of relativity was motivated by
examining how the electromagnetic laws transform in different
inertial frames \cite{einstein1905}.
As Einstein recalled in his autobiography, he was considering a Gedanken
experiment of chasing light:
If one could travel at the light velocity, the electromagnetic wave would
become a static field configuration, which is incompatible with the
Maxwell equations \cite{einstein1951, norton2004}.
Nevertheless, relativity is often taught in {\it Mechanics} in
{\it General Physics} \cite{kittle1973,feynman2011}.
The derivation of Lorentz transformation of space-time coordinates
is typically based on two postulates: I) The relativity
principle and II) the invariance of the light velocity.
The former is a generalization of the Galilean transformation
from mechanical laws to all physical laws including
electromagnetic ones, which is natural.
Nevertheless, although the light velocity invariance origins from
the invariance of the Maxwell equations, it looks mysterious and
unnatural.

Significant progresses have appeared in deriving relativity without
light in literature \cite{levyleblond1976,mermin1984,singh1986,pelissetto2015}.
Without loss of generality, consider the space-time transformation
in 1+1 dimensions.
Based on assumptions such as the homogeneity,
smoothness, and isotropy of space time, it can be derived
that only the Lorentz, Galilean, and rotation-like transformations
are possible.
They correspond to the hyperbolic, parabolic, and elliptic subgroups
of SL$(2,R)$, respectively.
Only the Lorentz and Galilean transformations meet the requirement
of causality.
If instantaneous interactions are further abandoned,
then Lorentz transformation is the only choice.

There have also been signficant efforts in elucidating the
intimate relation between relativity and electromagnetism
in general physics instructions.
The excellent textbook by E. M. Purcell, i.e., {\it Berkely Physics
Course, Vol II}, formulated magnetism as the relativistic consequence
of electricity.
Impressively, the magnetic field given by Ampere's law is consistent
with the result by applying the inertial frame transformation
to an electrostatic field as it should be.
Since magnetism is our daily life experience, this is
convincing for beginning students to accept relativity heartily.

In this article, we provide a new pedagogical way to re-derive
relativistic transformations via elementary electromagnetism
without the Maxwell equations.
Even the simple phenomenon of the drift velocity in a crossed
electric and magnetic field configuration, which is typically
a high school textbook problem, is incompatible with the
Galilean space-time transformation.
The postulate of light velocity invariance is not taken for
granted, but is derived as the upper limit of
drift velocities.
Simply by examining the transformations of drift velocities
in inertial reference frames, the transformation
laws of electromagnetic fields and the addition
law of velocities appear naturally, which are in contradiction
to the Galilean transform.
This would stimulate one to re-examine the space-time
coordinate transformation, which yields the
Lorentz transformation.
This is a logically natural way to start learning relativistic
physics.

\begin{figure}\centering
\includegraphics[width=0.8\linewidth]{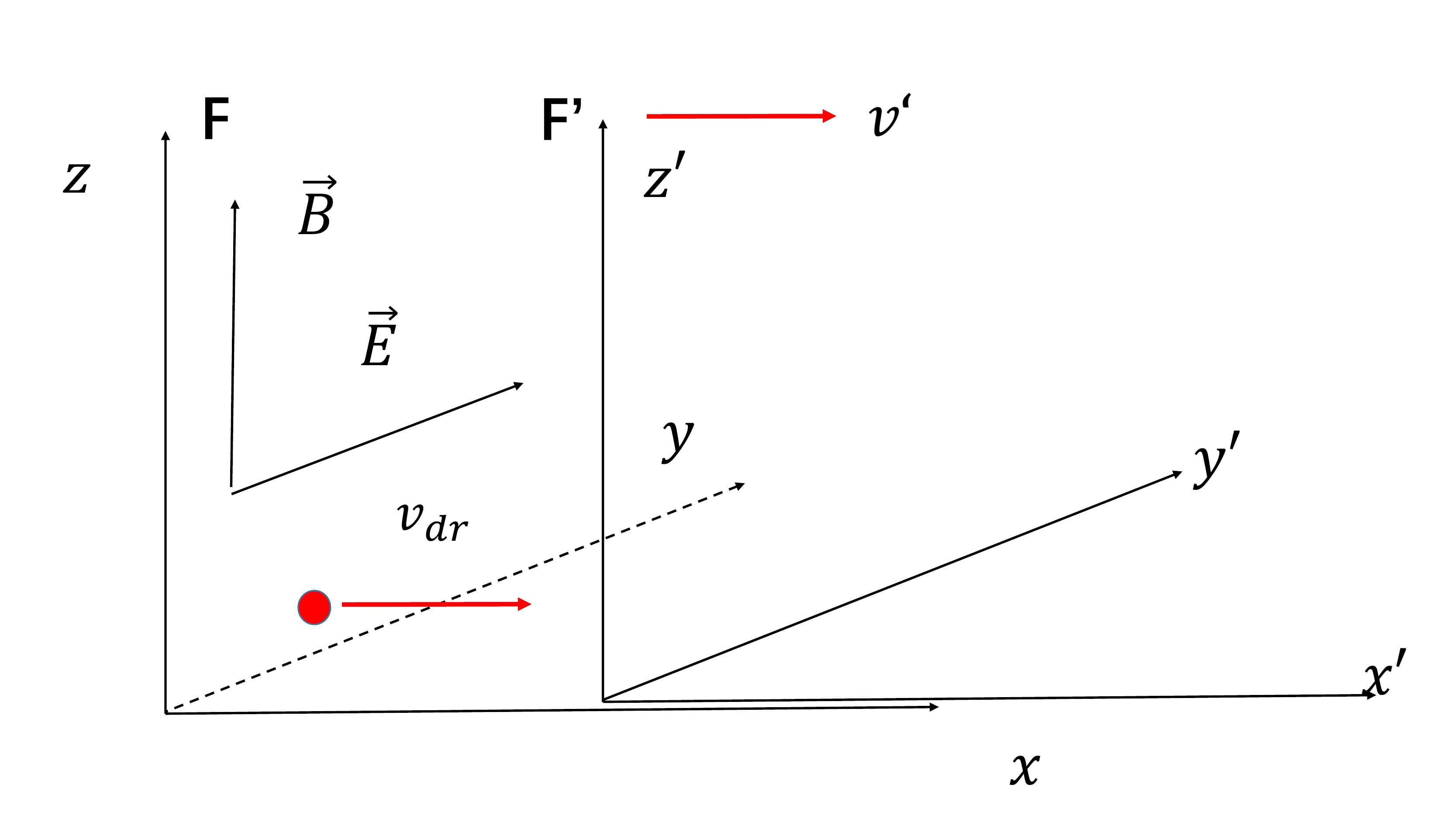}
\caption{Transformations of a crossed electric and
magnetic field configuration.
In the frame $F$, the magnetic field $\mathbf{B}$ is along the $z$-direction,
and the electric field $\mathbf{E}$ is along the $y$-direction.
$v_{dr}$ along the $x$-direction is the drift velocity of a charge
($v_q$) or a monopole ($v_g$) defined
in Eq. \ref{eq:q-drift} and Eq. \ref{eq:m-drift},
respectively.
The frame $F^\prime$ is moving relative to frame $F$ at the
velocity of $v^\prime$ along the $x$-direction.
}
\label{fig:frame}
\end{figure}

We warm up by reviewing elementary electromagnetism.
Consider a crossed field configuration in an inertial
reference frame $F$ as shown in Fig. \ref{fig:frame}.
Without loss of generality, it is assumed that
\bea
\mathbf{B}=B_z \hat z, \ \, \ \, \mathbf{E}=E_y \hat y.
\eea
For a charged particle $q$, its electric force and magnetic force
(Lorentz force) are expressed by
\bea
\mathbf{F}_q=q \left(\mathbf{E}+\frac{\mathbf{v}}{c} \times
\mathbf {B}\right).
\label{eq:charge}
\eea
The theory of electromagnetism actually allows the existence of
magnetic monopoles \cite{dirac1931,schwinger1968},
although it remains elusive in experimental detections
\cite{cabrera1982}.
For a monopole carrying a magnetic charge $g$,
its magnetic force and electric force are expressed \cite{jackson1998}
by
\bea
\mathbf{F}_g=g \left(\mathbf{B}-\frac{\mathbf{v}}{c} \times
\mathbf {E}\right),
\label{eq:monopole}
\eea
where the electric force becomes the Lorentz one.
Compared to Eq. \ref{eq:charge}, the electric Lorentz force exhibits an
opposite sign, whose physical meaning is
explained in Appendix \ref{sect:lorentz}.

We use the Gaussian unit enjoying the advantage
that electricity and magnetism are formulated in a symmetric way.
If in the SI unit, the force formulae are
\bea
\mathbf{F}_q&=&q \left(\mathbf{E}+\mathbf{v} \times
\mathbf {B}\right), \nn \\
\mathbf{F}_g &=&g \left(\mathbf{B}- \mu_0\epsilon_0 \mathbf{v} \times
\mathbf {E}\right)
= g \left(\mathbf{B}- \frac{ \mathbf{v}}{c^2} \times
\mathbf {E}\right).
\eea
So far, $c$ is just a quantity carrying the unit of velocity.

Based on Eq. \ref{eq:charge} and Eq. \ref{eq:monopole}, there exist
drift velocities.
For the charge $q$, its charge drift velocity is
\bea
\frac{v_q}{c}=\frac{E_y}{B_z},
\label{eq:q-drift}
\eea
at which the Lorentz force balances the electric force such that
the charged $q$ moves in a uniform velocity along
a straight line.
Similarly, the monopole drift velocity for the magnetic
charge $g$ is
\bea
\frac{v_g}{c}=\frac{B_z}{E_y}.
\label{eq:m-drift}
\eea

We define that a velocity $v$ is ``physical" if it can be taken
as the drift velocity for a charge or a monopole in a physically
realizable crossed electromagnetic field configuration.
Otherwise, it is ``unphysical".
When a charge or a monopole takes a physical drift velocity,
a co-moving frame can be defined in which the charge or monopole
is at rest.
Since velocity starts from zero continuously, when a drift velocity approaches zero, it should be physical without question.
We assume that if a velocity $v$ is already physical, any
velocity $u$
satisfying $|u|\le |v|$ is also physical.

We show that the at least one of charge drift velocity $v_q$
(Eq. \ref{eq:q-drift}) and monopole drift velocity $v_g$
(Eq. \ref{eq:m-drift}) is unphysical.
Contradiction would appear if both were physical.
Consider the co-moving frame $F_{dr}$ with the charge drift velocity $v_q$.
In such a frame, the charge is at rest, hence, both the magnetic
force and the electric force should vanish.
This means that the electric field is zero but the magnetic
field does not.
Otherwise, if both of them are zero in $F_{dr}$, they should
remain zero in the lab frame.
Then bring a magnetic charge $g$ in $F_{dr}$.
It would undergo acceleration, which cannot be removed by any inertial frame
transformation.
On the other hand, in the co-moving frame $F^\prime_{dr}$ with
a monopole at the drift velocity $v_g$, the monopole would be at
rest, showing the contradiction.

Since velocity $v$ is continuous, the physical and unphysical
regions should be separated by a threshold value $v_{th}$,
such that $v$ is physical if $|v|<v_{th}$ and unphysical
if $|v|>v_{th}$.
It is easy to prove that $v_{th}\le c$.
If it were not the case, i.e, $v_{th}> c$,
a set of values of $E_y$ and $B_z$ could be taken such that
$1<|B_z/E_y|< v_{th}/c$.
A charge would move with the drift velocity $v_q$, and
a monopole would move with the drift velocity $v_g$, and
both would be physical, which is impossible as shown before.

Let us check the limit of $|E|\ll |B|$.
The charge drift velocity $v_q\ll c$, which is physical.
In contrast, $v_g\gg c\ge v_{th}$, hence, the magnetic drift
velocity is unphysical.
Similarly, in the limit of $|E|\gg |B|$, the magnetic drift
velocity is physical, while the electric one is unphysical.

We can prove that it is impossible for $v_{th}< c$ either.
Otherwise, $v_{th}$ is non-universal, i.e.,
it takes different values in different inertial frames.
This would be in contradiction to the relativity principle,
since its different values could be used to distinguish
inertial frames.
This can be done by checking the transformation of $v_{th}$
in different frames (Eq. \ref{eq:addition}),
which will be derived later.

As a preparation, we derive the transformations of electric and
magnetic fields between different inertial frames.
The key idea of reasoning would be based on the relativity
principle -- all inertial frames are equivalent.
A physical drift motion is a uniform motion along a straight line,
hence, any inertial frame transformation does not change
this nature.
In other words, if the electric and magnetic forces are
balanced in one frame, they are balanced in any other
reference frames.

Without loss of generality, the frame $F^\prime$ is assumed moving with
a velocity $v^\prime$ along the $x$-axis with respect to the frame $F$.
These two frames share an O(2) symmetry, i.e, the rotation symmetry
with respect to the boost axis, i.e., the $x$-axis, and the reflection
symmetry with respect to any plane perpendicular to the $yz$-one.
The longitudinal components of the electric and magnetic fields, i.e.,
$E_x$ and $B_x$, are rotationally invariant.
But they transform differently under the mirror reflection with respect
to the $xz$-plane.
Hence, each of them transform to itself without mixing.
After boosting, $E_x$ can only change by a factor $\lambda$.
On the other hand, this factor should be independent of the boosting
direction.
If we perform one boost transformation, and then reverse the
boost back to the original frame,  $\lambda^2=1$ is arrived.
Since a boost can start with an infinitesimal velocity,
\bea
\lambda=1.
\label{eq:lambda}
\eea
A similar reasoning can be applied to $B_x$.
Hence, the longitudinal components of $\mathbf{E}$ and $\mathbf{B}$
are invariant
\bea
E_x^\prime=E_x, ~~ B_x^\prime=B_x.
\eea

As for the transverse components, $B_z$ and $E_y$
are odd under the reflection with respect to the $xz$-plane,
and even under the reflection with respect to the $xy$-plane.
In contrast, $B_y$ and $E_z$ transform oppositely under mirror
reflections with respect to these two planes.
Hence, $B_z$ and $E_y$ transform into each other under the
Lorentz boost, so does $B_y$ and $E_z$.

Now we assume the transformation between $E_y$ and $B_z$ as follows
\bea
\left( \begin{array}{c}
E_y^\prime\\
B_z^\prime
\end{array}
\right)=
\left( \begin{array}{cc}
a&b\\
c&d
\end{array}
\right)
\left(\begin{array}{c}
E_y \\
B_z
\end{array}
\right),
\label{eq:EBtransf}
\eea
where the matrix elements only depend on the boost velocity
$v^\prime$.
We choose a configuration that $E_y/B_z= v^\prime/c=\beta^\prime$ in the $F$
frame, such that the charge drift velocity $v_q=v^\prime$.
In the $F^\prime$ frame, the electric field should vanish, i.e.,
$E_y^\prime=0$, then
\bea
\frac{b}{a}=-\frac{E_y}{B_z}=-\beta^\prime.
\eea
Similarly, a configuration that $B_z/E_y=\beta^\prime$ is chosen
in the $F$ frame.
In this case, the $F^\prime$ frame is just the co-moving frame of a
magnetic charge with the drift velocity $v_g=v^\prime$.
Consequently $B_z^\prime=0$ in the $F^\prime$, yielding that
\bea
\frac{c}{d} =-\frac{B_z}{E_y}=-\beta^\prime.
\eea

Now let us consider a general charge drift velocity $v$ in the $F$-frame
with the corresponding electric and magnetic fields $B_z$ and $E_y$
such that $\beta=v/c=E_y/B_z$.
Under a Lorentz boost with velocity $\beta^\prime=v^\prime/c$, the
charge remains a drift motion.
The drift velocity $u$ in frame $F^\prime$ is
\bea
\frac{u}{c}= \frac{E_y^\prime}{B_z^\prime}
=\frac{a E_y +b B_z}{c E_y +d B_z}
=\frac{a}{d} \frac{\beta-\beta^\prime}{1-\beta\beta^\prime}.
\label{eq:u1}
\eea

Similarly, we can also prepare a monopole with the same drift
velocity $v$ in the $F$-frame, and choose the crossed field
configuration with $B_z/E_y=v/c=\beta$.
Again by the same Lorentz boost, the drift velocity becomes
$u$ in frame $F^\prime$, then
\bea
\frac{u}{c}=\frac{B_z^\prime}{E_y^\prime}
=\frac{c E_y +d B_z}{a E_y +b B_z}
=\frac{d}{a} \frac{\beta-\beta^\prime}{1-\beta\beta^\prime}.
\label{eq:u2}
\eea
Comparing Eq. \ref{eq:u1} with Eq. \ref{eq:u2}, $(a/d)^2=1$ is arrived.
Since $u/c=\beta$ in the limit of $\beta^\prime=0$,
we conclude that $a/d=1$.

So far we have derived the velocity addition law without employing
the space-time coordinate Lorentz transformation,
\bea
\frac{u}{c}=
\frac{\beta-\beta^\prime}{1-\beta\beta^\prime}.
\label{eq:vel-add}
\eea
It is remarkable that the above derivation only relies
on the formula of electric and magnetic forces
Eq. \ref{eq:charge} and Eq. \ref{eq:monopole},
and that a physical drift motion remains a physical drift motion
by inertial frame transformations.
These assumptions look more natural than the light velocity
invariance postulate.

Eq. \ref{eq:EBtransf} can be further refined by defining a quantity
$L=E_y^2-B_z^2$, which is known as the Lagrangian of the
electromagnetic field.
According to above information, we arrive at
\bea
L^\prime=E_y^{\prime 2}-B_z^{\prime 2}
=\lambda  L
\eea
where $\lambda=a^2 (1-\beta^{\prime 2})$.
$L$ should be insensitive to the boost direction since it
involves the square of fields, then $\lambda$ is an
even function of $\beta^\prime$.
By a similar reasoning in arriving at Eq. \ref{eq:lambda}, we also
conclude that $\lambda=1$, i.e., $a=\gamma=(1-\beta^2)^{-1/2}$.
Now we can refine the transform as
\bea
\left( \begin{array}{c}
E_y^\prime\\
B_z^\prime
\end{array}
\right)=
\left( \begin{array}{cc}
\gamma^\prime &-\gamma^\prime \beta^\prime\\
-\gamma^\prime \beta^\prime & \gamma^\prime
\end{array}
\right)
\left(\begin{array}{c}
E_y \\
B_z
\end{array}
\right).
\eea
By performing a rotation around the $x$-axis at $90^\circ$, we
arrive at
\bea
\left( \begin{array}{c}
E_z^\prime\\
B_y^\prime
\end{array}
\right)=
\left( \begin{array}{cc}
\gamma^\prime &\gamma^\prime \beta^\prime\\
\gamma^\prime \beta^\prime & \gamma^\prime
\end{array}
\right)
\left(\begin{array}{c}
E_z \\
B_y
\end{array}
\right).
\eea

With the above preparation, we prove that $v_{th}=c$ below.
In frame $F$, either $v_{th}$ is a realizable physical velocity,
or, as an upper limit of physical velocities.
After a Lorentz boost of velocity $v^\prime$, this threshold velocity
in frame $F^\prime$ becomes
\bea
\beta^\prime_{th}=\frac{v^\prime_{th}}{c}= \frac{\beta_{th}-\beta^\prime}
{1-\beta_{th}\beta^\prime}.
\label{eq:addition}
\eea
Due to the universality $v_{th}$, $\beta^{\prime}_{th}=\beta_{th}$,
then $\beta_{th}=1$, i.e., $v_{th}=c$.
We conclude that the light velocity is also invariant in any
inertial reference frame.

Now we are able to derive the Lorentz transformation of space-time
intervals.
By assuming
\bea
\left( \begin{array}{c}
\Delta x^\prime\\
c\Delta t^\prime
\end{array}
\right)=
\left( \begin{array}{cc}
a&b\\
c&d
\end{array}
\right)
\left(\begin{array}{c}
\Delta x \\
c\Delta t
\end{array}
\right),
\eea
the velocity addition law will be
\bea
\frac{u}{c}=\frac{\Delta x^\prime}{c \Delta t^\prime}
=\frac{a}{d} \frac{\frac{\Delta x}{c\Delta t} +\frac{b}{a}}{1+\frac{c}{d}\frac{\Delta x}{c\Delta t} }
=\frac{a}{d} \frac{\beta +\frac{b}{a}}{1+\frac{c}{d}\beta }
\label{eq:lorentz}.
\eea
Compared with Eq. \ref{eq:vel-add}, we conclude that
$a=d,  b/a=-\beta^\prime, c/d=-\beta^\prime$.

The length square of space-time interval  is defined as
$\Delta s^2=x^2-c^2\Delta t^2$.
According to light velocity invariance, if $\Delta s^2=0$,
then $\Delta s^{\prime,2}=0$.
Hence, for nonzero space-time interval, we could have
$\Delta s^{\prime,2}=\lambda \Delta s^2$, where $\lambda$
is a factor.
Again by the a similar reasoning in arriving at Eq. \ref{eq:lambda},
$\lambda=1$.
The Lorentz transformation is refined as
\bea
\left( \begin{array}{c}
\Delta x^\prime\\
c\Delta t^\prime
\end{array}
\right)=
\left( \begin{array}{cc}
\gamma^\prime & -\gamma^\prime \beta^\prime \\
-\gamma^\prime \beta^\prime & \gamma^\prime
\end{array}
\right)
\left(\begin{array}{c}
\Delta x \\
c\Delta t
\end{array}
\right).
\eea

In conclusion, we provide alternative derivations of the relativistic transformations of electromagnetic fields and the space-time coordinates.
By examining how the drift velocities of an electric charge and a
magnetic charge transform in different inertial frames, we arrive
at the same results as those in standard textbooks.
Nevertheless, the postulate of light velocity invariance is not
assumed, but actually is derived as the upper limit
of the physical drift velocities.
These results clearly show that the origin of relativity is deeply
rooted in electromagnetism.
Even for such an elementary electromagnetic phenomenon of
the drift velocity, it is inconsistent with the
Galilean space-time transformation and relativity is necessary.


\section{Author Declarations}
The authors have no conflicts to disclose.

\appendix
\section{Lorentz force for a magnetic monopole}
\label{sect:lorentz}
We justify the formula of Lorentz force for a magnetic monopole
as shown in Eq. \ref{eq:monopole}.
Consider a dyon system consisting of a charge $q$ and a monopole
$g$.
Consider the situation that the monopole $g$ is fixed at the
origin and the electron moves around.
The mechanical angular momentum $\mathbf{L}_{M,e}=\mathbf{r}\times m_e
\mathbf{v}$ is not conserved.
\bea
\frac{d}{dt} \mathbf{L}_{M,e}&=& \mathbf{r}\times m_e \frac{d \mathbf v}{dt}
=\mathbf{r}\times \left(\frac{e}{c} \mathbf{v}\times \mathbf{B}   \right)
\nn \\
&=& \frac{eg}{c} \frac{\mathbf{r}}{r^3}\times \left( \frac{d\mathbf{r}}{dt}
\times \mathbf{r} \right) \nn \\
&=& \frac{eg}{c} \frac{d}{dt} \left(\frac{\mathbf{r}}{r}\right).
\label{eq:OAM}
\eea
Hence, the right hand side can be viewed as as the time derivative
of the field contribution to angular momentum,
\bea
\mathbf{L}_{em}=-\frac{eg}{c} \left(\frac{\mathbf{r}}{r}\right).
\eea
In fact, it can be shown that
\bea
\mathbf{L}_{em}=\frac{1}{c}\int d^3\mathbf{r}~~ \mathbf{r}\times
\left( \mathbf{E}(\mathbf{r})\times \mathbf{B}(\mathbf{r}) \right).
\eea
Such that $\mathbf{L}_{tot}= \mathbf{L}_{M,e} +\mathbf{L}_{em}$
is conserved.

Now let us consider the situation that the electric charge is
fixed at the origin while the monopole is moving around.
Again, we need to define the total angular momentum as
\bea
\mathbf{L}^\prime_{tot}=L^\prime_{M,g}+\mathbf{L}^\prime_{em}.
\eea
Since $\mathbf{r}$ here is defined from the electric charge to
the monopole, which is opposite to the previous case, hence,
the field contribution of the angular momentum is
\bea
\mathbf{L}^\prime_{em}= \frac{eg}{c} \left( \frac{\mathbf {r}}{r}
\right).
\eea
Then the time derivative of the mechanical orbital angular
momentum should satisfy
\bea
\frac{d}{dt} \mathbf{L}_{M,g}=\mathbf{r}\times \mathbf{F}_{L,g}
=-\frac{d}{dt} \mathbf{L}^\prime_{em}=
-\frac{eg}{c} \frac{d}{dt} \left(\frac{\mathbf{r}}{r}\right).
\eea
Hence, the above result is consistent with Eq. \ref{eq:OAM}
if \bea
\mathbf{F}_{L,g}= -\frac{eg}{c} \mathbf{v}\times \frac{\mathbf{r}}{r^3}.
\eea
Hence,
\bea
\mathbf{F}_{L,g}= -g\frac{\mathbf{v}}{c}\times \mathbf{E}.
\eea


\end{document}